\documentclass{emulateapj}
\usepackage{apjfonts}

\newcommand{\te}{t_{\rm E}}

\newcommand{\thetae}{\theta_{\rm E}}
\newcommand{\zetavec}{\mbox{\boldmath $\zeta$}}

\shortauthors{HAN \& CHANG}
\shorttitle{LENSING SOURCE ELLIPTICITY}

\begin{document}

\title{Determination of Stellar Ellipticities in Future Microlensing Surveys}

\author{Cheongho Han}
\affil{Department of Physics, Institute for Basic Science
Research, Chungbuk National University, Chongju 361-763, Korea;\\
cheongho@astroph.chungbuk.ac.kr}

\and

\author{Heon-Young Chang}
\affil{Department of Astronomy and Atmospheric Sciences,
Kyungpook National University, 1370 Sankyuk-dong, Buk-gu, Daegu 702-701, 
Korea;\\ hyc@knu.ac.kr}


\begin{abstract}
We propose a method that can determine the ellipticities of source stars 
of microlensing events produced by binary lenses.  The method is based on 
the fact that the products of the caustic-crossing timescale, $\Delta t$, 
and the cosine of the caustic incidence angle of the source trajectory, 
$\kappa$, of the individual caustic crossings are different for events 
involving an elliptical source, while the products are the same for 
events associated with a circular source.  The product $\Delta t_\perp
=\Delta t \cos\kappa$ corresponds to the caustic-crossing timescale 
when the incidence angle of the source trajectory is $\kappa=0$.  For 
the unique determination of the source ellipticity, resolutions of 
at least three caustic crossings are required.  Although this requirement 
is difficult to achieve under the current observational setup based on 
alert/follow-up mode, it will be possible with the advent of future 
lensing experiments that will survey wide fields continuously at high 
cadence.  For typical Galactic bulge events, the difference in $\Delta 
t_\perp$ between caustic crossings is of the order of minutes depending 
on the source orientations and ellipticities.  Considering the monitoring 
frequency of the future lensing surveys of $\sim 6$ times/hr and the 
improved photometry especially of the proposed space-based survey, we 
predict that ellipticity determinations by the proposed method will be 
possible for a significant fraction of multiple caustic-crossing binary 
lens events involving source stars having non-negligible ellipticities.
\end{abstract}


\section{Introduction}

Microlensing was originally proposed to search for Galactic dark matter 
in the form of massive compact halo objects \citep{paczynski86}.  
However, with the ability to detect a large number of events, especially 
toward the Galactic bulge field, and to intensively monitor the detected 
events from follow-up observations, microlensing has been developed 
into a powerful tool to study various aspects of stellar astrophysics 
\citep{gould01}.  Especially, intensive observations of caustic-crossing 
binary lens events make it possible to measure the limb-darkening 
structure of stars, which in essence maps the atmospheric temperature 
as a function of depth.

Recently, \citet{rattenbury05} demonstrated an additional applicability 
of microlensing in stellar astrophysics by measuring the shape of the 
source star in microlensing event MOA 2002-BLG-33.  Due to rotation, 
stars are not exactly circular.  For example, Altair ($\alpha$ Aquillae) 
has an ellipticity of $e=(1-b^2/a^2)^{1/2}=0.23$ \citep{vanbelle01} and 
the Achernar ($\alpha$ Eridani), the flattest star ever measured, has an 
ellipticity of $e=0.59$ \citep{domiciano03}.  The star measured by 
\citet{rattenbury05} has an ellipticity of $e=0.2$.  Here $a$ and $b$ 
represent the angular sizes of the semi-major and semi-minor axes, 
respectively.  The nature of the source flattening makes the lensing 
lightcurve differ from that of a circular case.  However, detecting the 
deviation induced by the source flattening is very difficult because 
the deviation is much smaller than the deviation caused by limb darkening 
(see \S\ 3).  For the case of MOA 2002-BLG-33, an ellipticity measurement 
was possible because of the very special lens-source geometry in which
the source star was closely bounded on all sides by the caustics of the 
lens.  However, such a case is extremely rare.

In this paper, we propose a more general method that can determine 
the source ellipticity.  The method requires to resolve at least 
three caustic crossings and precisely measure the durations of the 
individual caustic crossings.  Although this requirement is difficult 
to achieve under the current observational setup based on 
alert/follow-up, it can be achieved by future lensing experiments 
that can survey wide fields continuously at high cadence.

The paper is organized as follows.  In \S\ 2, we describe the basics 
of caustic-crossing binary lens events involving elliptical source 
stars.  In \S\ 3, we demonstrate the difficulties in noticing the 
elliptical nature of source stars from the lightcurve of a single 
caustic crossing.  In \S\ 4, we explain how the source ellipticity 
is measured by the proposed method and demonstrate the feasibility 
of the method.  In \S\ 5, we summarize the results and conclude.

\section{Caustic Crossing of An Elliptical Source}

If a microlensing event is caused by a lens system composed of two 
masses, the resulting lightcurve can differ dramatically from the 
smooth and symmetric curve of a single lens event.  The main new 
feature of binary lensing is the caustics, which represent the 
positions in the source plane where the lensing magnification of 
a point source becomes infinite.  The set of caustics forms one or 
multiple closed figures where each figure is composed of concave line 
segments (fold caustics) that meet at points (cusps).  The number of 
separate sets of caustics is unity when the separation between the lens 
components is comparable to the Einstein ring radius $\thetae$, and 
it becomes two and three when the separation is substantially larger
and smaller than $\thetae$, respectively.  Due to the larger 
cross-section of the fold caustic compared to the cusp, majority 
of caustic crossing events are fold-caustic crossings.  Near a fold 
caustic, the magnification of the source star brightness is $A \propto 
u_\perp^{-1/2}$, where $u_\perp$ is the angular normal distance of the 
source from the fold caustic in units of the Einstein ring radius.  
Because of the divergent nature of the lensing magnification near the 
caustic, i.e.\ $A\rightarrow\infty$ as $u_\perp\rightarrow 0$, the 
lightcurve of a caustic-crossing event is characterized by sharp 
spikes.  Since the caustics form closed figures, caustic crossings 
occur in multiple times of two (at the entrance and exit of the caustic).  
\citet{mao91} predicted that $\sim 7\%$ of all Galatic microlensing 
events would be caustic-crossing events and this rate is roughly 
consistent with that detected by the MACHO \citep{alcock00} and OGLE 
\citep{jaroszynski02} lensing surveys.

The usefulness of caustic-crossing events lies in the divergent nature 
and large gradient of the lensing magnification near the caustic.  When 
a source approaches the caustic, different parts of the source star 
surface are magnified by different amounts (finite-source effect).  
Then, the lensing magnification of a finite source is equivalent to 
the intensity-weighted mean value averaged over the source star surface, 
i.e.
\begin{equation}
A_{fs}(\zetavec)=
{\int_S I(\zetavec') A(\zetavec+\zetavec') d \zetavec'
\over \int_S I(\zetavec')d \zetavec'},
\label{eq2.1}
\end{equation}
where $A$ denotes the point-source magnification, $\zetavec$ is the 
vector position of the center of the source, $\zetavec'$ is the 
displacement vector of a point on the source star surface with respect 
to the source star's center, $I(\zetavec')$ is the intensity profile 
of the source star, and the two-dimensional integral is over the 
source-star surface $S$.  Precise photometry during the caustic 
crossing enables one to measure the limb-darkening profiles of 
source stars \citep{witt95} and the limb-darkening coefficients of 
six stars with various stellar types were actually measure by this 
method \citep{albrow99a, albrow99b, albrow00, albrow01, afonso00, 
fields03, kubas05}.  It is also possible to study irregular structures 
on the source star surface, such as spots \citep{han00, chang02, hendry02}.

\begin{figure}[t]
\epsscale{1.15}
\plotone{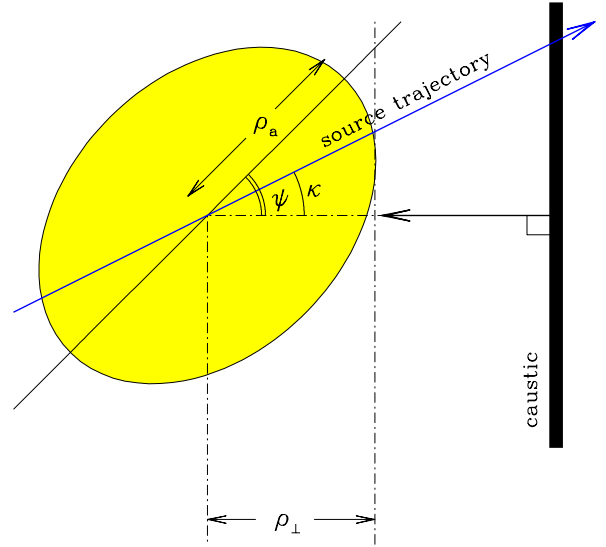}
\caption{\label{fig:one}
Definitions of the source orientation angle $\psi$, caustic incidence 
angle $\kappa$, normalized semi-major axis $\rho_a$, and the projected 
width $\rho_\perp$ of the source star.
}\end{figure}

The lensing magnification of a caustic-crossing binary lens event 
involving a {\it circular} source is characterized by eight parameters.  
Three of these are identical to those of a single point-lens event: the 
Einstein timescale, $t_{\rm E}$, the time of the closest source approach 
to the center of the binary lens system, $t_0$, and the separation between 
the source and center at that time, $u_0$.  For a binary lens, the lensing 
magnification is not radially symmetric and thus an additional parameter 
of the angle between the source trajectory and the binary axis, $\alpha$, 
is required to characterize the source trajectory.  Another two are the 
parameters related to the binary lens system: the mass ratio, $q$, and the 
separation (normalized by $\thetae$), $s$, between the lens components.  
The final two parameters are related to the source star: the normalized 
source radius, $\rho_\star=\theta_\star/\thetae$, and the limb-darkening 
coefficient.  Here $\theta_\star$ represents the angular radius of the 
source star.  For a limb-darkened source, the surface brightness profile 
is often modeled as 
\begin{equation}
{I(\theta) \over I(0)} =  
 1-\Gamma_\lambda \left[ 1- {3\over 2}
\sqrt{1-\left( {\theta\over \theta_\star} \right)^2}\right],
\label{eq2.2}
\end{equation}
where $\Gamma_\lambda$ is the limb-darkening parameter and $\theta$ is 
the angular distance measured on the source star surface from the center 
to the position within the source.  In the standard formalism, the linear 
limb darkening is expressed as 
\begin{equation}
{I(\theta) \over I(0)} =  
 1-c_\lambda \left[ 1- 
\sqrt{1-\left( {\theta\over \theta_\star} \right)^2}\right].
\label{eq2.3}
\end{equation}
Then, the relationship between the expressions of the linear limb-darkening 
coefficients is
\begin{equation}
c_\lambda = {3\Gamma_\lambda\over \Gamma_\lambda +2}.
\label{eq2.4}
\end{equation}
For an event involving an {\it elliptical} source, two additional 
parameters are needed.  These are the eccentricity of the source, 
$e$, and the orientation angle of the source major axis with respect 
to the perpendicular to the caustic, $\psi$  (see Fig.~\ref{fig:one}).  
Therefore, the total number of parameters required for the full 
characterization of a caustic-crossing binary lens event associated 
with an elliptical source is ten.

\section{Single Caustic Crossing}

Due to the intrinsic difference in shape, the lightcurve of a
caustic-crossing binary lens event involving an elliptical source 
differs from that of an event associated with a circular source.  
However, detecting the deviation induced by the source flattening is 
difficult.  This is in part because the flattening of the majority of 
stars is small.  However, we find that even in the case of severely 
flattened stars the deviation caused by the source flattening is 
substantially smaller than the deviation caused by the limb darkening.  
In this section, we demonstrate this fact.

\begin{figure}[t]
\epsscale{1.2}
\plotone{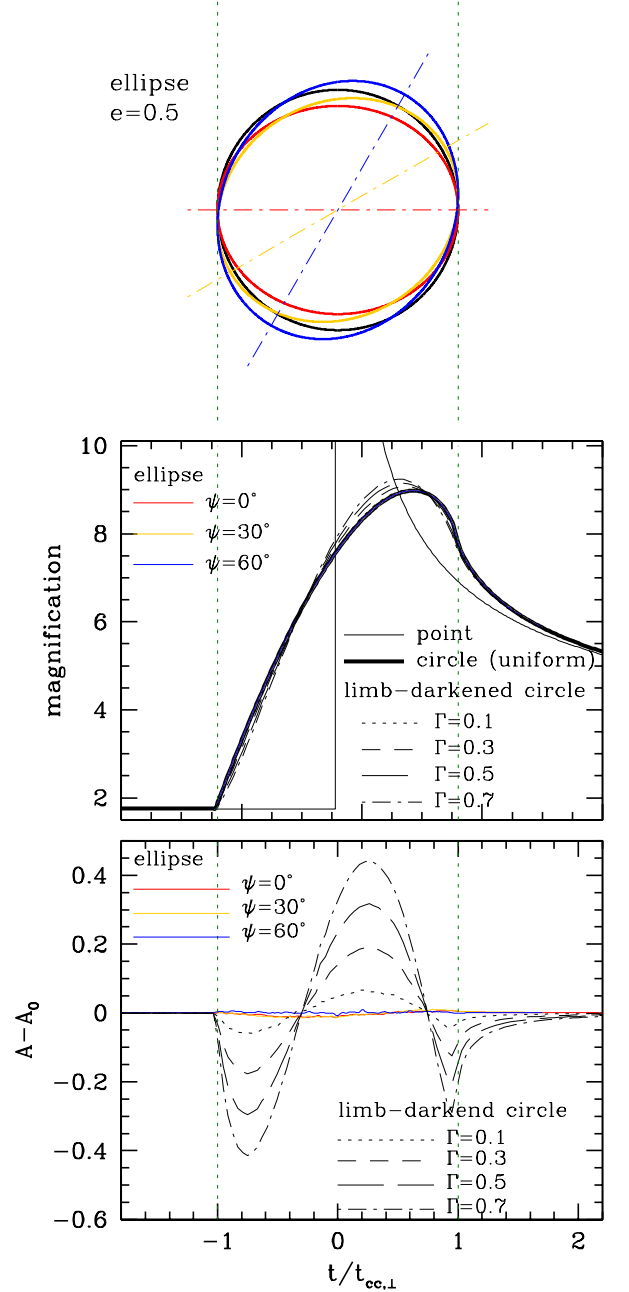}
\caption{\label{fig:two}
Lightcurves of caustic-crossing binary lens events associated with 
source stars having various shapes, sizes, and limb-darkening coefficients 
(middle panel) and the deviations from a uniform circular case (bottom 
panel).  Top panel shows the shapes and orientations of tested source 
stars.
}\end{figure}

To see the pattern of deviations induced by the source flattening and 
compare the magnitude of the deviation to that caused by limb darkening, 
we simulate events associated with source stars having various shapes, 
sizes, and limb-darkening coefficients.  For the cases of events with 
circular sources, we consider limb darkening.  We test four cases of 
limb darkening coefficients of $\Gamma_\lambda=0.1$, 0.3, 0.5, and 0.7, 
which cover most values ranging from main-sequence to giant stars.  
The tested value of the normalized source radius is $\rho_\star=0.02$, 
which roughly corresponds a giant source star of a typical Galactic 
bulge event.  For events with elliptical sources, however, we do not 
consider limb-darkening effect, so as to isolate the magnitude of 
deviation caused only by source flattening.  We test three different 
elliptical cases where the individual source stars have a common 
ellipticity but  different orientation angles.  For the ellipticity, 
we adopt a relatively large value of $e=0.5$ ($b/a=0.866$).  The tested 
values of the orientation angle are $\psi=0^\circ$, $30^\circ$, and 
$90^\circ$.  We set the normalized semi-major axes, $\rho_a=a/\thetae$, 
of the individual elliptical source stars with different orientation 
angles so that the times of the caustic entrance and exit (and thus 
the duration of caustic crossing) match those of the circular source 
case.  As a result, the individual elliptical source stars have different 
semi-major axes with ratios of $\rho_a(\psi=0^\circ) : \rho_a(\psi=
30^\circ) : \rho_a( \psi = 60^\circ) = 1.000 : 1.035 : 1.110$ (see the 
top panel of Fig.~\ref{fig:two}).  The duration of caustic crossing is 
proportional to the projected width of the source star in the direction 
perpendicular to the caustic, $\rho_\perp$ (projected width, see 
Fig.~\ref{fig:one}).  The projected width is related to the semi-major 
axis and orientation angle by 
\begin{equation}
\Delta \rho_\perp = \rho_a
\left[ (1-e^2)^2+\cot^2\psi \over 1-e^2+\cot^2\psi \right]^{1/2}.
\label{eq3.1}
\end{equation}
All lightcurves are produced by the same binary lens system with $q=0.8$ 
and $s=1.2$ and the parameters related to the source trajectory are 
$u_0=0.27$ and $\alpha=0^\circ$.  The angle of the source trajectory 
with respect to the perpendicular to the caustic, $\kappa$ (caustic 
incidence angle, see Fig.~\ref{fig:one}), is very close to zero.

In Figure~\ref{fig:two}, we present the results of the simulation where 
the middle panel shows the magnifications of the individual tested 
events, $A$, and the bottom panel shows the deviations from a uniform 
circular case $A_0$.  The top panel shows the shapes and orientations 
of the tested source star.  The lightcurves of the limb-darkened circular 
source cases show a characteristic pattern in which the deviation is 
negative when the caustic is located on the fainter edge of the source 
while the deviation is positive when the caustic is positioned on the 
brighter central part.  The elliptical cases show more complex pattern 
depending on the the size and orientation of the source star.  However, 
regardless of the variation, the deviation induced by the source 
flattening is substantially smaller than the deviation caused by the 
limb-darkening effect.

\section{Multiple Caustic Crossings}

\subsection{Basic Scheme of the Method}

Although difficult with only a single caustic-crossing lightcurve,
it is possible to notice the elliptical nature of the source star 
if both the entrance and exit of the caustic are well resolved.
This is possible because the products of the caustic-crossing timescale 
and the cosine of the caustic incidence angle, i.e.\ $\Delta t\cos\kappa$, 
of the individual caustic crossings are different each other for an 
elliptical source.   For a circular source, on the other hand, the 
products are always the same regardless of the source orientation.  
The product corresponds to the caustic-crossing timescale when the 
incidence angle of the source trajectory is $\kappa=0^\circ$ (hereafter 
we refer to the product as the `normal incidence caustic-crossing 
timescale' and denote as $\Delta t_\perp$).  The normal incidence 
caustic-crossing timescale is related to the source parameters by
\begin{equation}
\Delta t_\perp = \Delta t\cos\kappa = 2
\left[ (1-e^2)^2+\cot^2\psi \over 1-e^2+\cot^2\psi \right]^{1/2} 
\rho_a\te.
\label{eq4.1}
\end{equation}
We note that the caustic-crossing timescale $\Delta t$ is defined as 
the duration between the source's enter and exit of the caustic.
For each caustic crossing, the caustic-crossing timescale is measured 
from the lightcurve during the caustic crossing.  The incidence angles 
$\kappa$ of the individual caustic crossings and the Einstein timescale 
$\te$ are determined from the model fit to the overall shape of the 
lightcurve.

\begin{figure}[t]
\epsscale{1.1}
\plotone{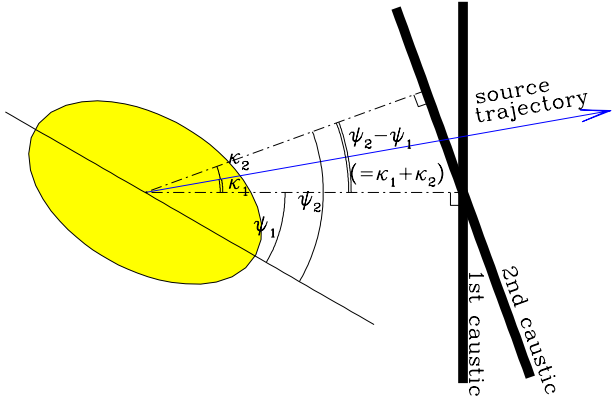}
\caption{\label{fig:three}
Relation between the source orientation angles $\psi_1$ and $\psi_2$
with respect to the perpendiculars to first and second caustics.
$\kappa_1$ and $\kappa_2$ are the incidence angles of the source star
to the first and second caustics, respectively.
}\end{figure}

Although the elliptical nature of the source star can be noticed from 
two caustic crossings, unique  determination of the source ellipticity 
requires an additional crossing.  The normal incidence caustic-crossing 
timescale is a function of three unknowns of $\rho_a$, $e$, and $\psi$, 
i.e.\ $\Delta t_\perp = \Delta t_\perp (\rho_a, e, \psi)$.  If one 
resolved the lightcurve profiles of two caustic crossings, the number 
of measured quantities increases to two, i.e.\ the normal incidence 
caustic-crossing timescales at the first and second caustic crossings, 
$\Delta t_{\perp,1}$ and $\Delta t_{\perp,2}$, but the number of 
unknowns remains the same.  This is because the source orientation 
angles of the individual caustic crossings are related each other.
This is demonstrated in Figure~\ref{fig:three}.  In the presented case, 
the orientation angle at the second caustic crossing $\psi_2$ is related 
to the orientation angle at the first crossing $\psi_1$ by $\psi_2 = 
\psi_1 + \kappa_1 + \kappa_2$, where $\kappa_1$ and $\kappa_2$ are the 
caustic incidence angles at the first and second caustic crossings, 
respectively.  However, with two caustic crossings, the number of unknowns 
still exceeds that of that of measured quantities.  Therefore, if one has 
an additional lightcurve of caustic crossing, the numbers of unknowns and 
measured quantities become the same, and thus one can uniquely determine 
all three unknowns of $\rho_a$, $e$, and $\psi_1$ (and subsequently 
$\psi_2$ and $\psi_3$).

There are two channels for multiple caustic crossings.  One is the 
case in which the source trajectory passes two separate sets of 
caustics.  The other channel is the case where the source trajectory 
is aligned closely along a fold of the caustic as shown in an example 
in Figure~\ref{fig:four}.  Multiple crossings through the second channel 
can occur because fold caustics have concave curvature.  If the 
individual caustic crossings of these events are resolved, the events 
will provide precious chances of measuring the source ellipticities.

\subsection{Feasibility of the Method}

The basic requirement of the proposed method is resolving multiple 
caustic crossings.  Currently, microlensing follow-up observation 
is possible only after an alert is issued.  Under this strategy, 
the first caustic crossing is usually missed due to the unpredictable 
nature and short duration of the caustic crossing.  The second caustic 
crossing may be resolved because the crossing can be roughly predicted 
from the characteristic ``U'' shape of the lightcurve when the source 
is inside the caustic \citep{jaroszynski01}.  In addition, there is a 
guarantee that the second crossing should occur and thus it is worth 
devoting observational resources for the resolution of the second 
caustic crossing.  However, there is no characteristic pattern 
between the pairs of the caustic crossings, and thus it would be 
difficult to predict when the additional crossings will occur.  
Furthermore, there is no guarantee that these additional crossings 
will occur.  Therefore, considering that the main object of the 
follow-up observation is detecting deviations induced by planetary 
companions, it would be difficult to devote the limited resources 
waiting for unpredictable additional caustic crossings.

However, the limitation can be overcome with the advent of future 
lensing experiments that can survey wide fields continuously at high 
cadence by using very large format cameras.  Two such surveys were 
proposed.  The {\it Galactic Exoplanet Survey Telescope} ({\it GEST}), 
whose concept was succeeded by the {\it Microlensing Planet Finder} 
({\it MPF}), is a space mission to be equipped with a 1.0m -- 1.5m 
telescope \citep{bennett02}.  The `Earth Hunter' project plans to 
achieve high monitoring frequency by using three 2m-class wide-field 
ground-based telescopes scattered over the southern hemisphere (A.\ 
Gould, private communication).  These next-generation surveys will 
dispense with the alert/follow-up mode and instead simultaneously 
obtain densely sampled lightcurves of all microlensing events in the 
field.  These experiments are expected to detect of order 5000 events 
per year.  Therefore, the set of multiple caustic-crossing events for 
which an ellipticity measurement is possible would not be vanishingly 
small.

\begin{figure}[t]
\epsscale{1.15}
\plotone{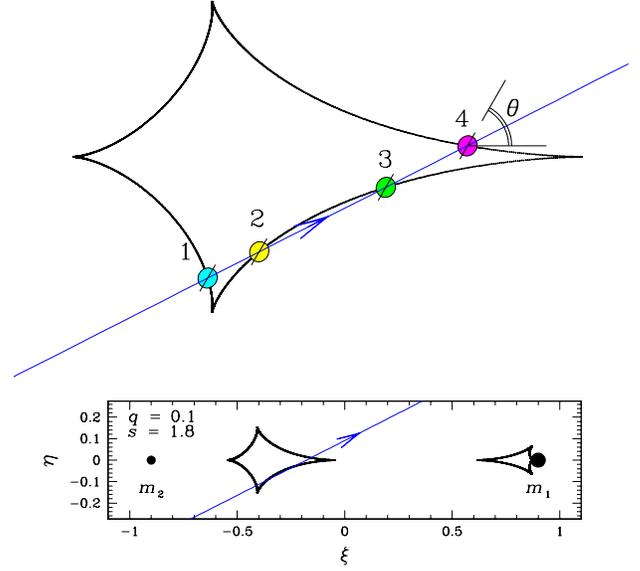}
\caption{\label{fig:four}
An example event with multiple caustic crossings.  The small ellipses
on the source trajectory (straight line with an arrow) represent the 
source star at the times of the caustic crossings, which are numbered 
according to the order of time.  The size and flattening of the source 
star are exaggerated to better show the orientations of the major axis.  
The angle $\theta$ is the orientation angle of the source star major 
axis with respect to the binary axis.  The small bottom panel shows 
the lens system geometry, where the positions of the lens components 
($m_1$ and $m_2$), caustics, and source trajectory are marked.
}\end{figure}

\begin{deluxetable*}{cccccccccc}
\tablecaption{Normal Incidence Caustic-crossing Timescales \label{table:one}}
\tablewidth{0pt}
\tablehead{
\multicolumn{1}{c}{$\theta$} &
\multicolumn{1}{c}{crossing} &
\multicolumn{1}{c}{$\kappa$} &
\multicolumn{1}{c}{$\psi$} &
\multicolumn{3}{c}{$\Delta t_\perp$ (hr)} &
\multicolumn{3}{c}{max$|\Delta t_{\perp,i}- \Delta t_{\perp,j}|$ (hr)} \\
\colhead{(${}^\circ$)} &
\colhead{number} &
\colhead{(${}^\circ$)} &
\colhead{(${}^\circ$)} &
\colhead{$e=0.2$} &
\colhead{$e=0.3$} &
\colhead{$e=0.4$} &
\colhead{$e=0.2$} &
\colhead{$e=0.3$} &
\colhead{$e=0.4$} \\
}
\startdata
30  & 1 & 9.718  & 12.718  & 1.9182  & 1.9162  & 1.9137  & 0.0359  & 0.0820  &  0.1488 \\
    & 2 & 77.689 & 80.689  & 1.8823  & 1.8342  & 1.7649  &         &         &         \\
    & 3 & 80.424 & 77.424  & 1.8831  & 1.8363  & 1.7691  &         &         &         \\
\smallskip
    & 4 & 53.208 & 50.208  & 1.8976  & 1.8703  & 1.8339  &         &         &         \\

60  & 1 & 9.718  & 42.718  & 1.9026  & 1.8819  & 1.8548  & 0.0293  & 0.0667  & 0.1204  \\
    & 2 & 77.689 & 69.311  & 1.8863  & 1.8438  & 1.7839  &         &         &         \\
    & 3 & 80.424 & 47.424  & 1.8995  & 1.8746  & 1.8418  &         &         &         \\
\smallskip
    & 4 & 53.208 & 20.208  & 1.9156  & 1.9105  & 1.9042  &         &         &         \\

90  & 1 & 9.718  & 72.718  & 1.8848  & 1.8403  & 1.7769  & 0.0341  & 0.0775  & 0.1393  \\
    & 2 & 77.689 & 39.311  & 1.9049  & 1.8870  & 1.8638  &         &         &         \\
    & 3 & 80.424 & 17.424  & 1.9167  & 1.9129  & 1.9082  &         &         &         \\
    & 4 & 53.208 & 9.792   & 1.9189  & 1.9177  & 1.9162  &         &         &         \\
\enddata
\end{deluxetable*}

To investigate the feasibility of implementing the proposed method in 
future lensing surveys, we estimate the magnitude of the difference 
in the normal incidence caustic-crossing timescales between caustic 
crossings, $\Delta t_{\perp,i}-\Delta t_{\perp, j}$, expected from 
typical Galactic binary lens events.  As an example, we choose an event 
produced by a binary lens system with a mass ratio and separation 
of $q=0.1$ and $s=1.8$, respectively.  Figure~\ref{fig:four} shows 
the caustics and source trajectory of the example event.  For the 
source star, we assume a normalized semi-major axis of $\rho_a=0.002$, 
which corresponds to that of a F0 main-sequence source star of a 
typical Galactic bulge event produced by a low-mass stellar lens 
with distances to the lens and source of $D_L\sim 6$ kpc and $D_S\sim 8$ 
kpc.  We test three different ellipticities of $e=0.2$, 0.3, and 0.4, 
which correspond to the axis ratios of $b/a=0.979$, 0.954, and 0.917, 
respectively.  We test three cases of source orientations where the 
angles between the source major axis and the binary axis are $\theta=
30^\circ$, $60^\circ$, and $90^\circ$ (see the definition of $\theta$ 
in Fig.~\ref{fig:four}).  The small ellipses on the trajectory in 
Figure~\ref{fig:four} represent the source star at the times of caustic 
crossings.  We designate the individual crossings by numbers according 
to the order of time.  We note that the source star in the figure is
bigger and flatter than the actual tested stars to better show the 
orientations of their major axes.

The results are summarized in Table~\ref{table:one}, where we present 
the crossing number, caustic incidence angle $\kappa$, source 
orientation angle $\psi$, and the normal incidence caustic-crossing 
timescales $\Delta t_\perp$ of the individual caustic crossings.
Also presented are the maximum values of the differences between the 
normal incidence caustic-crossing timescales, i.e.\ ${\rm max}|\Delta 
t_{\perp,i}-\Delta t_{\perp, j}|$.  The range of the difference is 
order of minutes depending on the source orientation and ellipticity.

Then, what would be the precision of $\Delta t$ measurement?  In an 
ideal case where there are $N$ equally spaced measurements over the 
caustic crossing with a fractional photometric precision $\sigma$, the 
times of the caustic entrance and exit can be known with a precision of 
$\sim \sigma\Delta t/\sqrt{N}$.  In reality, however, the actual error,
$\delta(\Delta t)$ would be larger than this due to various factors.  
Estimating error in $\Delta t$ measurement considering all these factors 
is difficult.  We, therefore, estimate the ratio of the actual error to 
the value of the ideal case, i.e. $f=\delta(\Delta t)/(\sigma\Delta t/
\sqrt{N})$, based on the error estimate in actually observed caustic 
crossings.  For this, we choose two events with well resolved caustic 
crossings of MACHO 98-SMC-1 \citep{albrow99b} and OGLE-1999-BUL-23 
\citep{albrow01}.  Assuming $\sigma=2\%$, we find that $f\sim 2.7$ and 
1.9 for the individual events.  By taking $f=2.0$ as a representative 
value, we then estimate the precision of $\Delta t$ measurement in the 
next-generation experiments as
\begin{equation}
\delta(\Delta t)\sim f {\sigma \Delta t\over \sqrt{N}}.
\label{eq4.2}
\end{equation}
By assuming a typical caustic crossing timescale of $\Delta t\sim 2$ 
hrs, the photometric precision of $\sigma\sim 0.5\%$, and the monitoring 
frequency of 6 times/hr (and thus $N\sim 12$), we find that $\delta 
(\Delta t)\sim 0.3$ min.  Considering that the typical value of the 
difference $|\Delta t_{\perp,i}-\Delta t_{\perp,j}|$ is order of 
minutes, we predict that ellipticity measurements will be possible 
for a significant fraction of multiple caustic-crossing events involving 
source stars having non-negligible ellipticities.

\section{Conclusion}

We have proposed a method that can determine the ellipticities of 
source stars of microlensing events produced by binary lenses.  The 
method is based on the fact that the normal incidence caustic-crossing 
timescales of the individual caustic crossings are different for events 
involving an elliptical source, while the timescales are the same for 
events associated with a circular source.   For the unique determination 
of the source ellipticity, resolutions of three and more caustic crossings 
are required.  Although this requirement is difficult to achieve under 
the current observational setup, it will be possible with the advent of 
future lensing experiments that will survey wide fields continuously at 
high cadence.  For typical Galactic bulge events, the difference in 
$\Delta t_\perp$ between caustic crossings is order of minutes.  
Considering the high monitoring frequency of the future lensing surveys 
and the improved photometry especially of the proposed space-based survey, 
we predict that ellipticity determinations by the proposed method will be 
possible for a non-negligible number of events.

\acknowledgments
We would like to thank A.\ Gould for making helpful comments.
This work was supported by the Astrophysical Research Center for the
Structure and Evolution of the Cosmos (ARCSEC) of the Korea Science \&
Engineering Foundation (KOSEF) through the Science Research Program (SRC)
program.

\end{document}